# Information Exchange, Meaning and Redundancy Generation in

# Anticipatory Systems: Self-organization of Expectations - the case of Covid-19

Inga A. Ivanova [1]

## Abstract

When studying the evolution of complex systems one refers to model representations comprising various descriptive parameters. There is hardly research where system evolution is described on the base of information flows in system. The paper focuses on the link between the dynamics of information and system evolution. Information, exchanged between different system's parts, before being processed is first provided with meaning by the system. Meanings are generated from the perspective of hindsight, i.e. against the arrow of time. The same information can be differently interpreted by different system's parts (i.e. provided with different meanings) so that the number of options for possible system development is proliferated. Some options eventually turn into observable system states. So that system evolutionary dynamics can be considered as due to information processing within the system. This process ia considered here in a model representation. The model under study is Triple Helix (TH) model, which was earlier used to describe interactions between university, industry and government, to foster innovations. In TH model the system is comprised of three interacting parts where each part process information in a different way. The model is not limited to the sphere of innovation and can be used in a broader perspective. I conceptualize TH model in the framework of three compartment model used to describe infectious diseases. The paper demonstrates how the dynamics of information and meaning can be incorporated in the description of Covid-19 infection outbreak. The results show correspondence of model predictions with observable infection dynamics.

---
[1] Institute for Statistical Studies and Economics of Knowledge, National Research University Higher School of Economics (NRU HSE), 20 Myasnitskaya St., Moscow, 101000, Russia; inga.iva@mail.ru * corresponding author

### 1. Introduction

Complex biological and social systems can be considered as set of interacting agents (actors or sub-systems). Actors are connected with each other via communications. Actors, as nodes, and communications, as links, form structural network. Structural network carry system evolution on the top of structural layer in terms of functions, such as supply, demand and control (Leydesdorff, Ivanova & Meyer, 2019). Interaction between two actors occasionally generates change from the system's previous state. When the system comprises three or more actors each third actor disturbs the communication between other two actors. This disturbance can reinforce or constrain this change. Thus autocatalytic (reinforcing) or stabilizing (constraining) cycles may arise. Mixed structures in addition to cycles are also possible (Fig.1).

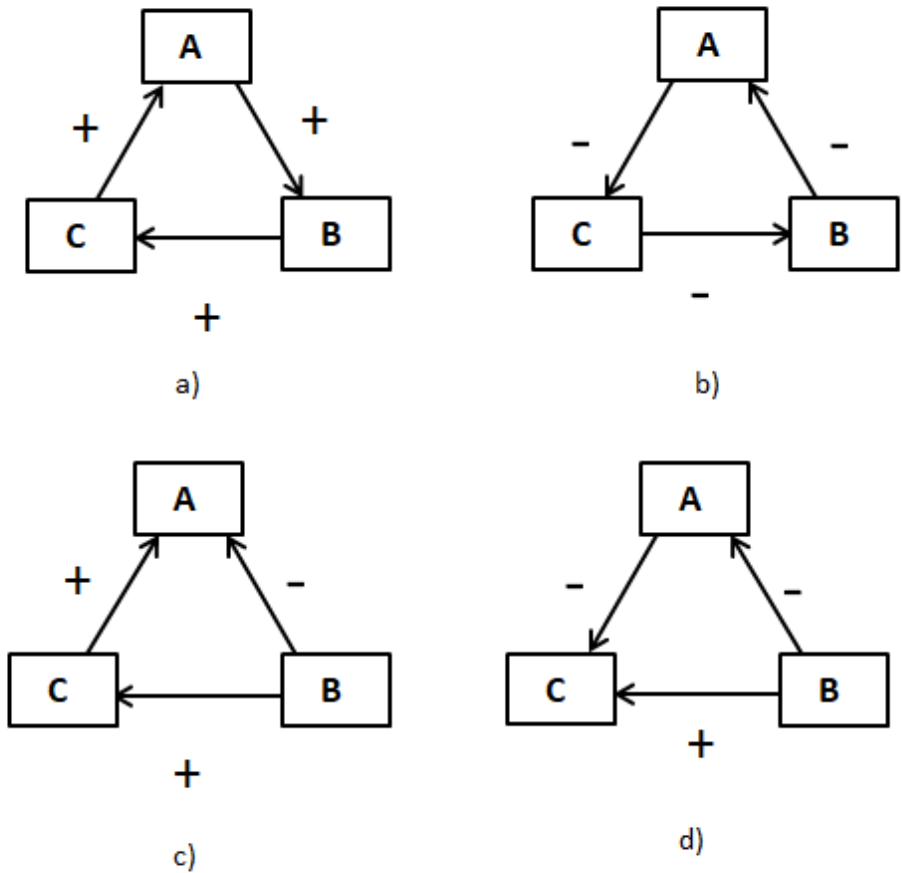

Fig. 1 Schematic representation of possible relations in three-component network: a) autocatalytic cycle; b) stabilizing cycle; c) and d) mix of autocatalytic and stabilizing links

In stabilizing mode the system keeps to its previous historical states generated with arrow of time and evolves along the trajectory. While in autocatalytic mode the system relies on available options, which have not yet been realized and self-organizes. In a mixed mode the system temporary deviates from historical trajectory.

When the actors are of different nature they also span a correlation network. Correlation means that actors are different or similar only to some degree with respect to processing the information. In other words they are positionally differentiated. Latent structures organize different meanings into structural components. These structures are driven by coding rules (Leydesdorff, 2010). That is, the same information may be supplied with different meanings by different actors. E.g. the information about some technology may be treated differently with respect to its scientific importance, industrial applicability or patentability. The structural differences among the coding and decoding algorithms provide a source of additional options in reflexive and anticipatory communications, meaning generating structures act as selection environments (Leydesdorff, 2021).

Meanings produce expectations about possible system states which are generated with respect to future moments (i.e. against the arrow of time). Expectations can be considered as options. The more options possess the system, the more probable the system is to change when self-organization is present.

An example of a system comprising positionally differentiated actors is the Triple Helix of university-industry-government relations (Etzkowitz and Leydesdorff, 1998). A TH is expected to reveal non-linear interactions among the helices since it is intrinsically non-linear system. Non-linear interactions mean that system behavior is described by non-linear equations.

Ivanova, Leydesdorf (2014a) showed that the origin of non-linearity lies in TH triadic structure. TH actors can be mapped as components of three dimensional vector and their interactions presented as rotations of this vector. Non-linearity stems from non-commutativity of vector rotation the in a three dimensional space.

Actors not only exchange information, but can also share meanings provided by partially overlapping perspectives, Partial overlaps generate redundancy since the same information can be provided with different meanings and meanings are used to generate new redundant options. This redundancy can be considered as the number of new and not yet realized options in a system of innovations. Redundancy can be measured quantitatively using the synergy indicator (Leydesdorff, Dolfsma, & van der Panne, 2006; Leydesdorff, Ivanova, 2014).

The synergy indicator measures the generation of redundancy as a result of a three-way interaction among TH actors (Leydesdorff, Ivanova, & Meyer, 2019). TH also incorporates anticipatory dynamics based on expectations. The theory and computation of anticipatory systems can help numerically evaluate and predict the incursive and recursive dynamics in a model representation (Rosen, 1985; Dubois, 1998).

The measurement of expectations with synergy indicator allows estimate the synergy in interaction among the actors which in some way can help to predicts system evolution. However two major issues that require further research can be distinguished: 1) redundancy is evaluated as a static measure without direct reference to system's evolutionary dynamics and indeed one should develop a dynamical equivalent to TH indicator (Leydesdorff, 1991), also interactions in complex systems tend to involve circularities which can not be described by static products of probabilities (Krippendorff, 2009); and 2) what way redundancy is linked to the transformation of the system? The answer to these questions can open up additional opportunities in studying TH like systems of different origin, including economic, biological, and social ones. The relations between evolutionary theory and systems theory can further be specified using communication theory (Theil, 1972).

The first research question of the present paper is how TH redundancy dynamics arises from interactive relationships among historical and anticipated states? I argue that the dynamics of redundancy in these interactions is subject to self-organization provided by non-linear mechanisms, run with non-linear evolutionary differential equation. This dynamics manifests itself in recognizable longitudinal patterns. The patterns in turn can be used to predict future system evolution and are linked to logistic function.

As a second research question I study whether there is a link between generated redundancy and TH-like system transformation. I argue that model predictions are in good agreement with Covid-19 epidemic spread data. The obtained results show that the unfoldment of Covid-19 infection matches model predictions in terms of successive infection wave amplitudes ratios. This suggests a link between redundancy dynamics and system transformation dynamics.

## 2.      Method

In a TH model the communication between each of two selection environments is shaped by the third selection environment (Sun & Negishi, 2010). This mechanism drives the system transformation and is known as "triadic closure" (Granovetter, 1973; Bianconi *et al*.,  2014) Two cycles with positive and negative feedback and feedforward loops are possible: autocatalytic and stabilizing the dynamic in organizational formats one. A positive cycle amplifies a change from previous system state and can be considered as system self-organization, while negative cycle corrects this change and stabilizes the system along a historical trajectory so that system evolution is driven by two opposing tendencies (Ulanowicz, 2009). These cycles can be modeled as two three dimensional vectors $Q$ and $P$ which rotate in opposite directions (Ivanova & Leydesdorff, 2014b). Redundancy $R$ is the result of a balance between historical stabilization and self-organization (Ivanova & Leydesdorff, 2014a):

$$R \sim P^2 - Q^2 \qquad (1)$$

Also mutual redundancy $R_{123}$ in a TH system can be measured with help of synergy indicator (Ulanowicz, 1986; 2009; Leydesdorff, Dolfsma, & van der Panne, 2006):

$$R_{123} = H_1 + H_2 + H_3 - H_{12} - H_{13} - H_{23} + H_{123} \qquad (2)$$

where $H$ is an entropy of one, two and three variables distribution

$$H_{i..j} = - \sum_{i..j} p_{i..j} log_2 p_{i..j} \qquad (3)$$

$p_{i..j}$ – corresponding probabilities.

A calculus of redundancy is a complement to Shannon's calculus of information (Bar-Hillel, 1955). TH redundancy can be either positive or negative, depending on the nature of actor's interaction. Negative redundancy is the result of self-organization in the communications which provides more opportunities and positive redundancy indicates the historical organization in the instantiations by exploiting existing opportunities.

The first term in Eq. (1) can be considered as due to historical organization (which adds to positive redundancy) and the second term corresponds to self-organization and which augments negative redundancy. Historical organization relates to historically realized options which are generated via recursive mode and self-organization bears on new, not yet realized options, generated via incursive mode[2]. Comparing two expressions for redundancy provided by Eqs.(1) and (2) one can identify positive and negative terms in Eq.(1) with positive and negative terms in Eq.(2).

Redundancy cannot infinitely grow or decrease. It was shown by Ivanova and Leydesdorff (2014b) that in a TH model redundancy time evolution is cyclical. For temporal cyclic systems probabilities $p_i$ can oscillate around their average values $p_{i0}$ in harmonic or non-harmonic mode.

For non-harmonic oscillations, one can derive (see Appendix A):

$$\frac{1}{k}\frac{d^2 p_i}{dt^2} = -(p_i - p_{i0}) + \alpha(p_i - p_{i0})^2 + C_i \qquad (4)$$

The probability density function $P$ is the solution of a non-linear evolutionary equation:

$$P_T + PP_X + \delta P_{XXX} + C_1 = 0$$

and can be written in the form of a solitary wave, otherwise named as soliton (see Appendix B for the derivation):

$$P = 2\kappa^2 ch^{-2}\left[\kappa\left(X - 4\kappa^2 T + \frac{C_1}{2}T^2\right)\right] - C_1 T \qquad (5)$$

Or in a more general form:

---

[2] Recursive systems use their past states to modulate the present ones, while hype- incursive systems employ future anticipated states to shape their present states (e.g. Leydesdorff & Dubois, 2004)

$$P = n(n+1)\kappa^2 ch^{-2}\left[\kappa\left(X - 4\kappa^2 T + \frac{c_1}{2}T^2\right)\right] - C_1 T \qquad (6)$$

An impulse in the form as in Eq. 6 eventually evolves in a train of *n* solitary waves with amplitudes $2\kappa^2$, $8\kappa^2$, $18\kappa^2$ …$2n^2\kappa^2$ and the corresponding velocities $4\kappa^2$, $16\kappa^2$, $32\kappa^2$, … $4n^2\kappa^{2\,3}$ (Miura, 1976 ).

Additional term at the right hand side of Eq.5 attenuates the initial impulses amplitudes with time. Wave, described by Eq. 5, moves forward and after time span: $T_1 = {8k^2}/{C_1}$ returns to the origin. There is a relation between solitons amplitude ratios and time span ratios: ${T_2}/{T_1} = {A_2}/{A_1}$.

Moving frame represent the arrow of time. Information obtained via informational exchange is processed with communication codes and expectations with respect to future time are generated at a system's level.  These expectations can be considered as redundancy density presenting non-realized but possible options, distributed along time interval, which can, not in all but in some cases, trigger subsequent actions. Here expectations are analytical events (options) and actions are historical events[4] which can be observed after some time span with respect to expectations (as if expectations move against the arrow of time and turn into actions). In other words, there is a dynamic of the actions in historical events at the bottom and a dynamic of expectations at the upper level operating reflexively. Expectations are eventually transformed into actions representing new system state[5]. The moments of time when expectations turn to actions correspond to the moments when the impulse returns to the origin.

---

[3] Eq.12 refers to a net soliton solution. In case of arbitrary initial perturbation it evolves in a train of solitons moving off to the right and oscillatory dispersive state moving off to the left [33].

[4] Shannon (1948) defined the proportion of non-realized but possible options as redundancy, and the proportion of realized options as the relative uncertainty or information.

One can obtain the probability function from the probability density function according the formula:

$$p(t) = \int_{-\infty}^{t} P(t)dt = th(t) + 1 = {2e^{2t}}/{1 + e^{2t}} \qquad (7)$$

The resulting probability function is a sigmoid curve which is frequently used to describe the behavior of a dynamic system. Often this behavior resembles a series of logistic waves. Growth and diffusion patterns can be decomposed in one or more sigmoid curves.

### 3.      Results

*Modeling Covid-19 pandemic*

One of simple models depicting infectious disease dynamics is the *SIR* model originally introduced by Kermack and McKendrick (1927). The model comprises dynamic interactions among three groups (compartments): *S* – the part of population which is susceptible to infection, *I* – the part of population which has been infected, *R* – the part of population which has been removed from the first two parts because they are recovered, deceased or vaccinated. Values *S*, *I*, *R* vary over time. The three compartment members have different criteria regarding infection that governs their behavior. In this respect compartments can be considered positionally

---

[5] According the Second law of thermodynamics system's entropy increase with time

differentiated which reminds TH model comprising three positionally differentiated institutional spheres. Three compartments operation can be formulated in terms of supply, demand and control functions. E.g. for medical innovations supply, demand and control can be articulated in terms of new treatments, innovation in terms of diseases, and practical experiences and evaluations (Nelson et al., 2011). Here demand is expressed in morbidity rate, supply in susceptibility to infection and control – in infection resistance.

The three compartment of the SIR model can be sequentially reduced to Verhulst (logistic) equation with the parameters determined by the basic characteristic of epidemic process which provide an accurate description of the Covid-19 epidemic when infection outbreak can be fitted by single logistic curve (e.g. Postnikov, 2020). Though the spread of the disease can rarely be fitted by one curve and often resemble a succession of overlapping curves, so that the overall logistic behavior is hard to discern and analyze. Daily statistics of new cases can provide visual guidance to distinguishing successive epidemic waves.

Here I analyze a few examples of Covid-19 spread in USA, Canada Russia, UK, Belgium, Finland, Japan and Israel. The data for analysis were downloaded from: https://public.tableau.com/profile/kffdata#!/vizhome/KFF-CoronavirusTracker-Update4-10-20/Overview and https://www.ecdc.europa.eu/en/publications-data/downloadtodays-data-geographic-distribution-covid-19-cases-world. Instead of summary cases and new cases I consider the probability densities as representations of the chances of infection. Probabilities are defined as a ratio of a summary cases a certain point in time to country's population. Probability densities are ratios of daily new cases to country's population.

Daily data were used to distinguish different periods presented by infection surge. Data in each period were processed with MathCad software to determine the parameters of the logistic curve:

$$F_i(t) = {A_i}/{(1 + B_i \exp(-C_i t)} \tag{9}$$

Finally these curves were merged to approximate the whole dataset.

$$F(t) = \sum_i (f_i(t + \Delta t_i) + d_i)_i \tag{10}$$

$\Delta t_i$- horizontal shift; $d_i$- vertical shift. Differentiating Eq. (10) by $t$ one obtains daily cases fit:

$$f(t) = \sum_i A_i C_i sech^2 \left( C_i t + \Delta t_i - \ln \left( {B_i}/{2} \right) \right) \tag{11}$$

The function $F(t)$ determines parameters of empirically observed infection waves, such as amplitude, width and position. These parameters are further compared with model predicted ones.

Descriptive statistics comprising parameters of the data fitting (Figs 2a-9a) is provided in Table 1. Infection outspread is country-specific and unfolds in two, three or four waves. $A_i$, $B_i$, $C_i$ –parameters of logistic curves (Eq. 9), $T_i$ – time shifts.

Table 1 Parameters for the data fitting for USA, Canada, Russia, UK, Belgium, Finland, Japan and Israel

| | USA | Canada | Russia | UK | Belgium | Finland | Japan | Israel |
|---|---|---|---|---|---|---|---|---|
| $A_1$ | $5.054*10^{-3}$ | $2.89*10^{-3}$ | $6.443*10^{-3}$ | $4.307*10^{-3}$ | $5.08*10^{-3}$ | $1.297*10^{-3}$ | $1.339*10^{-4}$ | $1.921*10^{-3}$ |
| $B_1$ | $3.296*10^3$ | 82.91 | 84.426 | 13.884 | 82.824 | 3186.664 | $5.973*10^4$ | 443.886 |
| $C_1$ | 0.086 | 0.064 | 0.045 | 0.067 | 0.105 | 0.076 | 0.119 | 0.142 |
| $T_1$ | 86 | 69 | 98 | 49 | 35 | 97 | 58 | 38 |
| $A_2$ | 0.015 | 0.022 | 0.024 | 0.024 | 0.052 | $7.336*10^{-3}$ | $5.334*10^{-4}$ | $9.961*10^{-3}$ |
| $B_2$ | 28.948 | 138.206 | 41.625 | 411.247 | $0.06*10^5$ | 313.688 | 165.534 | 504.644 |
| $C_2$ | 0.054 | 0.034 | 0.033 | 0.062 | 0.065 | 0.038 | 0.074 | 0.086 |
| $T_2$ | 176 | 302 | 268 | 240 | 228 | 320 | 174 | 143 |
| $A_3$ | 0.08 | - | - | 0.038 | - | 0.01 | $3.012*10^{-3}$ | 0.027 |
| $B_3$ | 22.432 | - | - | 30.289 | - | 24.599 | 89.729 | 45.026 |
| $C_3$ | 0.032 | - | - | 0.086 | - | 0.048 | 0.048 | 0.099 |
| $T_3$ | 316 | - | - | 308 | - | 417 | 313 | 203 |
| $A_4$ | - | - | - | - | - | - | - | 0.061 |
| $B_4$ | - | - | - | - | - | - | - | 99.469 |
| $C_4$ | - | - | - | - | - | - | - | 0.059 |
| $T_4$ | - | - | - | - | - | - | - | 319 |

Below are the set of plots illustrating how the data are fitted. Figs. 2 -9 shows Logistic curve fitting to the graph of Covid-19 infection for USA, Canada Russia, UK, Belgium, and Finland, Japan and Israel respectively in relation to cumulative chance of infection and new cases chance of infection.

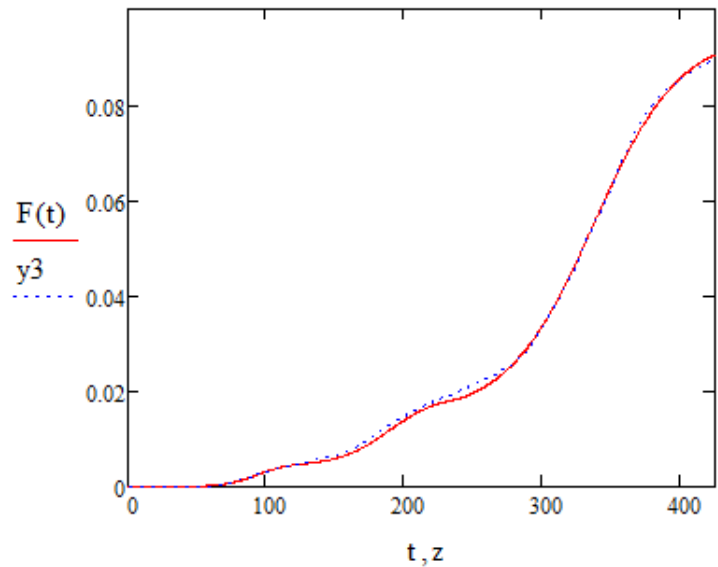

Fig.2a. Logistic curve fitting (solid line) of USA Covid-19 chance of infection data (x axis: days since the first case, y axis: dot line - empirical data, solid line – function fit).

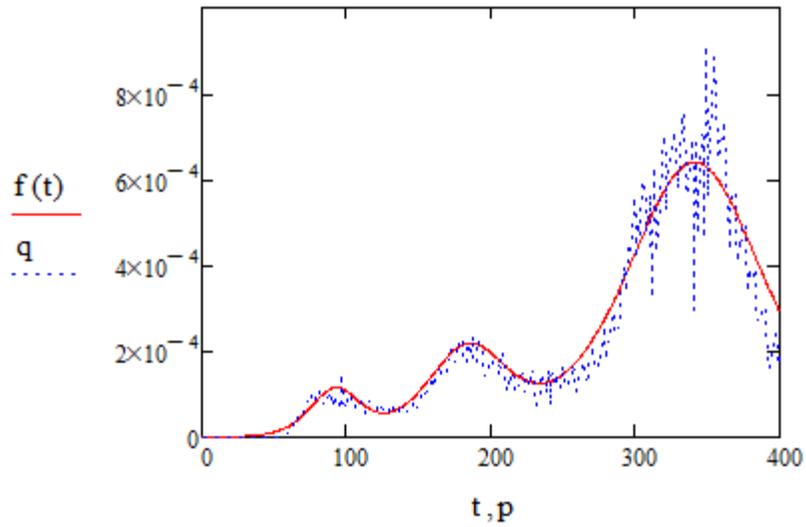

Fig.2b. USA Covid-19 new cases chance of infection three wave approximation (x axis: days since the first case, y axis: dot line - empirical data, solid line – function fit)

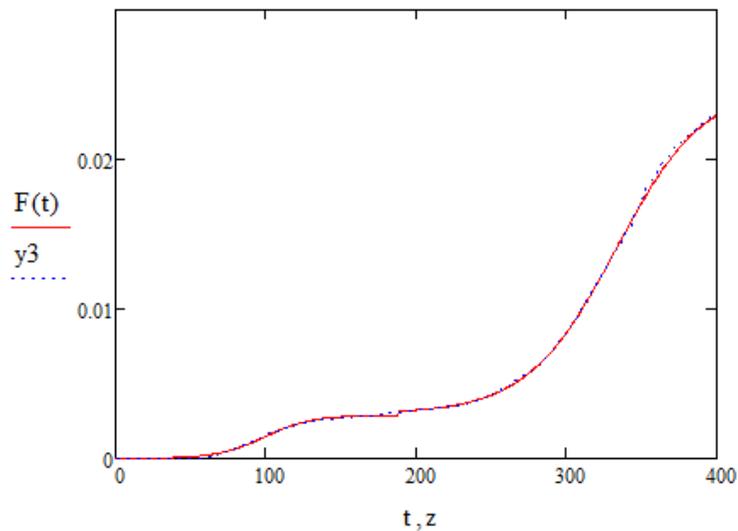

Fig.3a. Logistic curve fitting (solid line) of Canada Covid-19 chance of infection (x axis: days since the first case, y axis: dot line - empirical data, solid line – function fit)

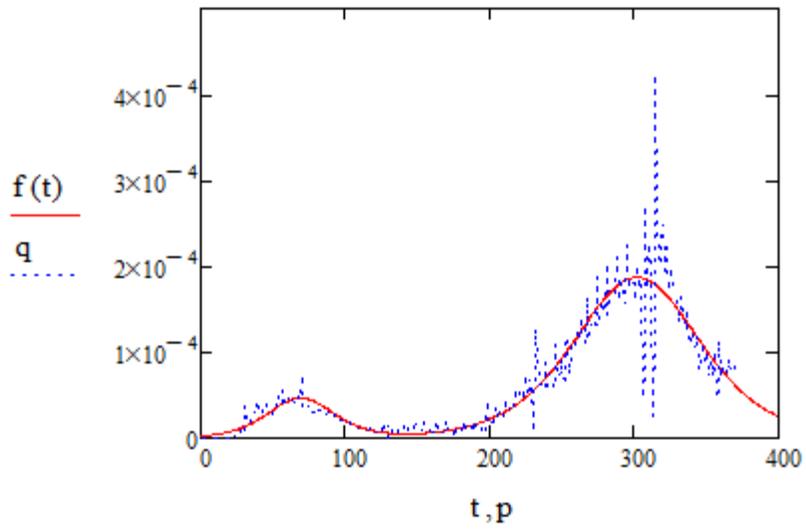

Fig.3b. Canada Covid-19 new cases chance of infection two wave approximation (x axis: days since the first case, y axis: dot line - empirical data, solid line – function fit)

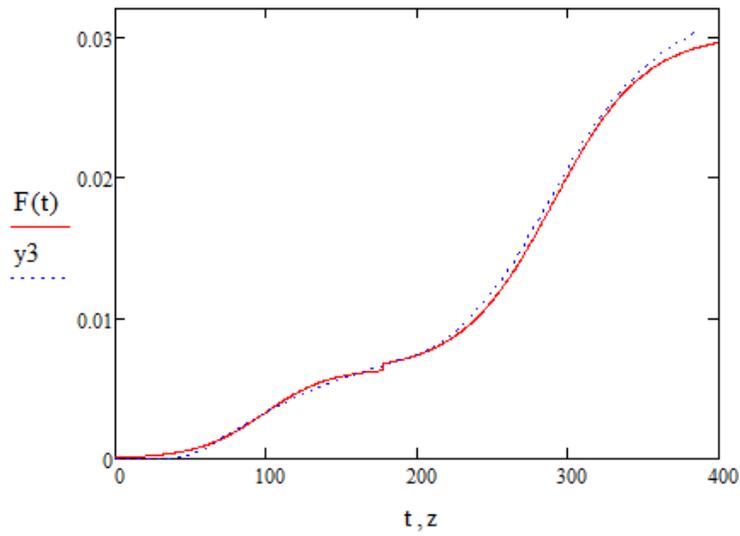

Fig.4a. Logistic curve fitting (solid line) of Russia Covid-19 chance of infection logistic curve approximation (x axis: days since the first case, y axis: dot line - empirical data, solid line – function fit)

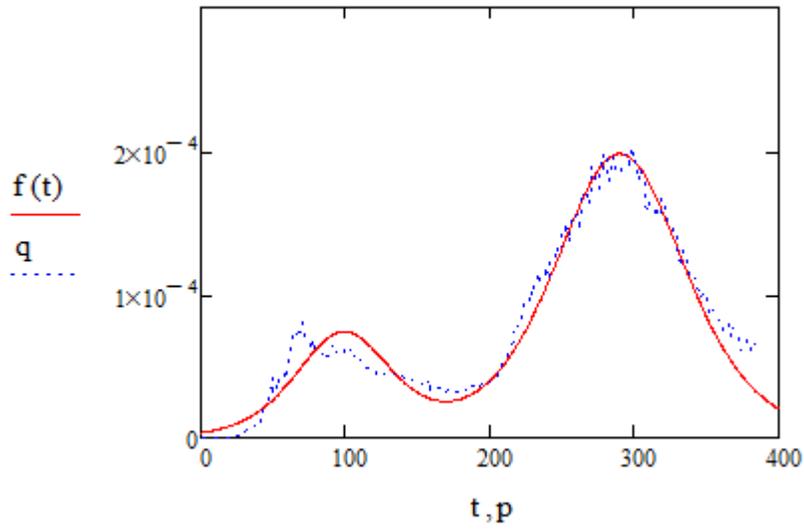

Fig.4b. Russia Covid-19 new cases chance of infection two wave approximation (x axis: days since the first case, y axis: dot line - empirical data, solid line – function fit)

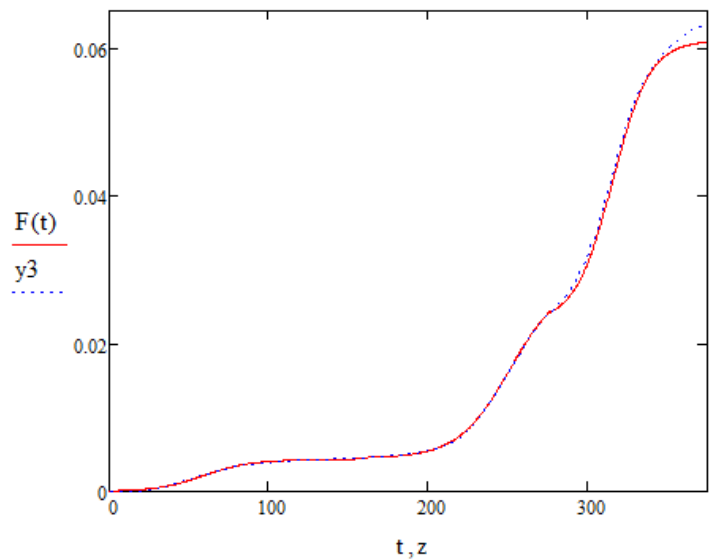

Fig.5a. Logistic curve fitting of UK Covid-19 chance of infection (x axis: days since the first case, y axis: dot line - empirical data, solid line – function fit

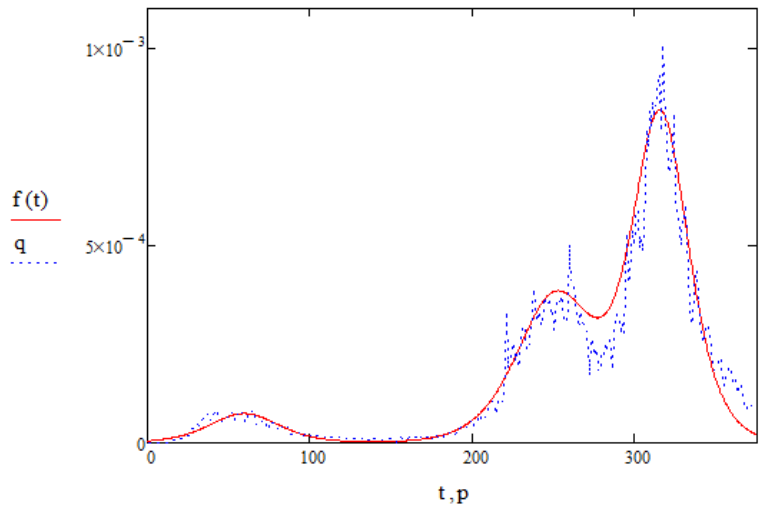

Fig.5b. UK Covid-19 new cases three wave approximation (x axis: days since the first case, y axis: dot line - empirical data, solid line – function fit)

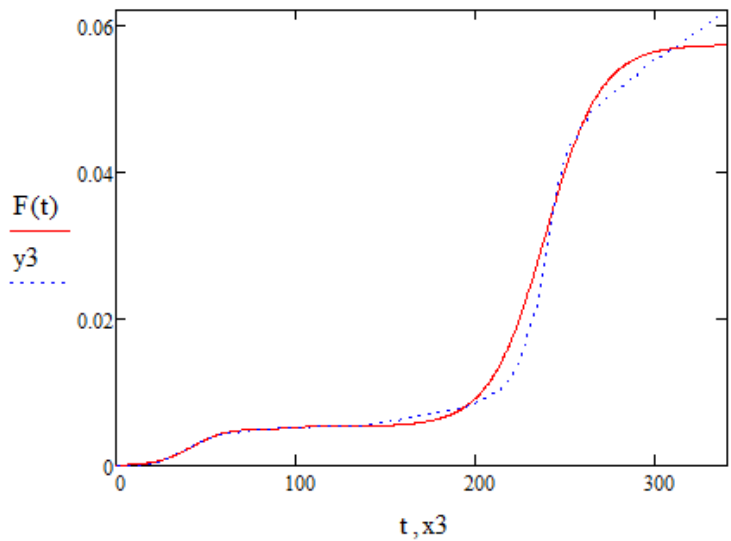

Fig. 6a Logistic curve fitting of Belgium Covid-19 daily infection data (x axis: days since the first case, y axis: dot line - empirical data, solid line – function fit)

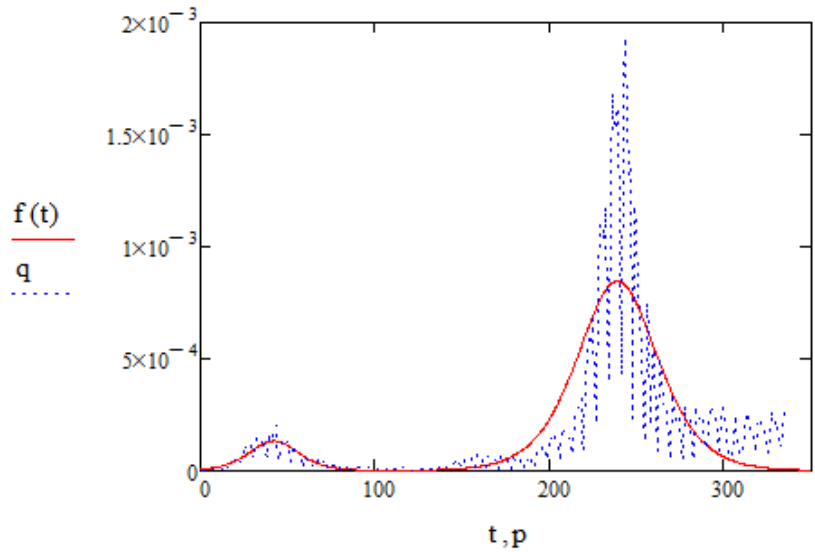

Fig.6b. Belgium Covid-19 new cases chance of infection two wave approximation (x axis: days since the first case, y axis: dot line - empirical data, solid line – function fit)

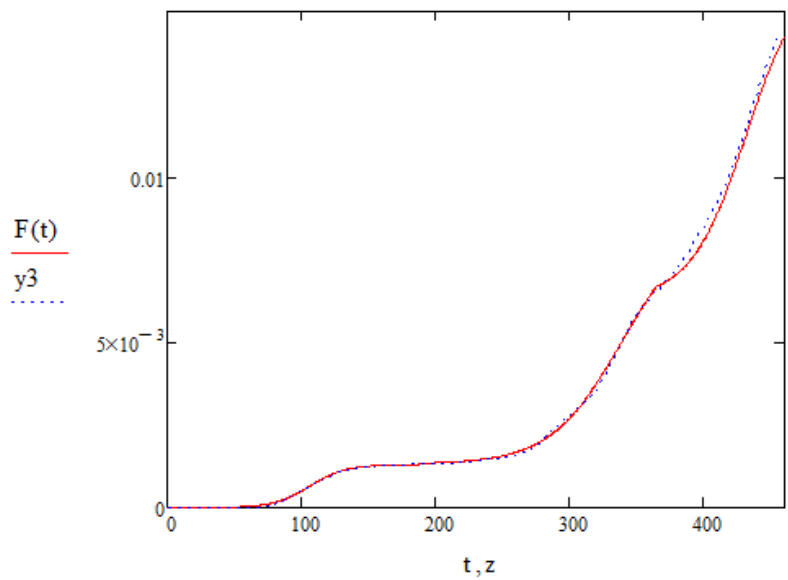

Fig.7a. Logistic curve fitting of Finland Covid-19 chance of infection (x axis: days since the first case, y axis: dot line - empirical data, solid line – function fit)

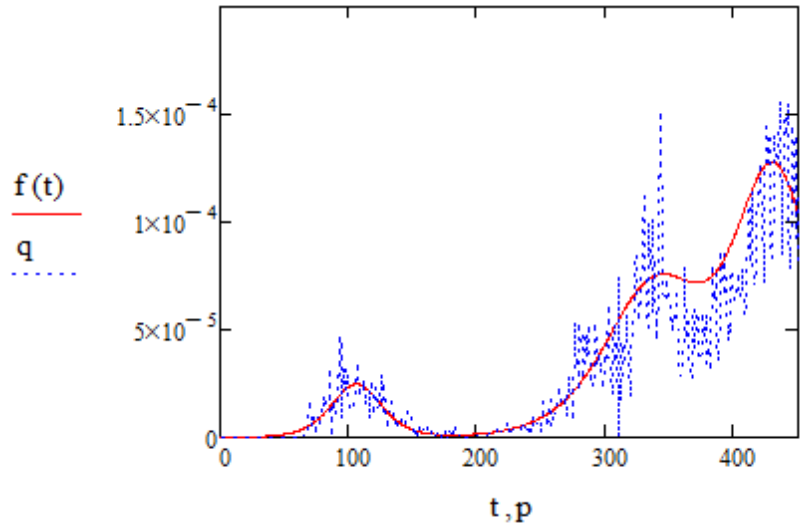

Fig.7b. Finland Covid-19 new cases chance of infection three wave approximation (x axis: days since the first case, y axis: dot line - empirical data, solid line – function fit)

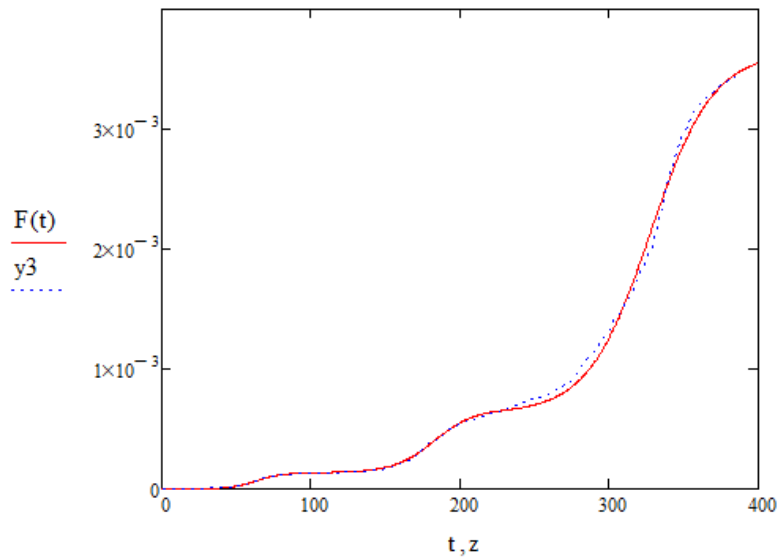

Fig.8a. Logistic curve fitting of Japan Covid-19 chance of infection (x axis: days since the first case, y axis: dot line - empirical data, solid line – function fit)

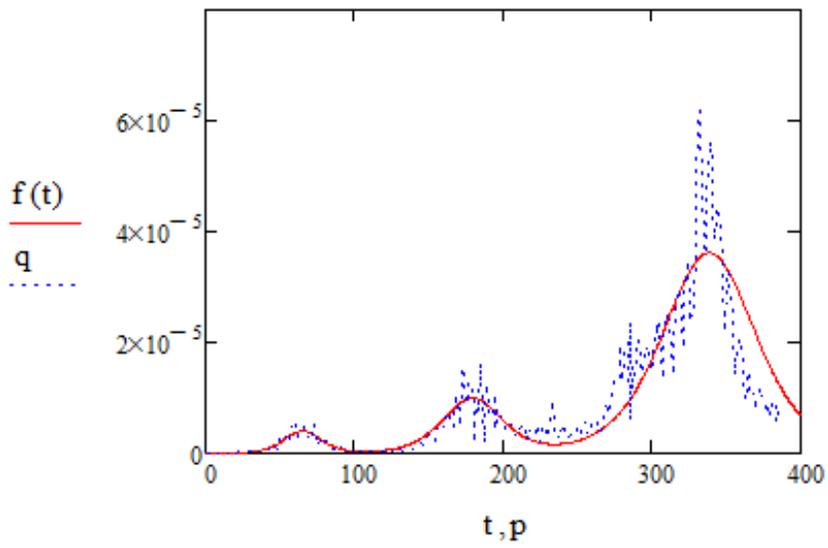

Fig.8b. Japan Covid-19 new cases chance of infection three wave approximation (x axis: days since the first case, y axis: dot line - empirical data, solid line – function fit)

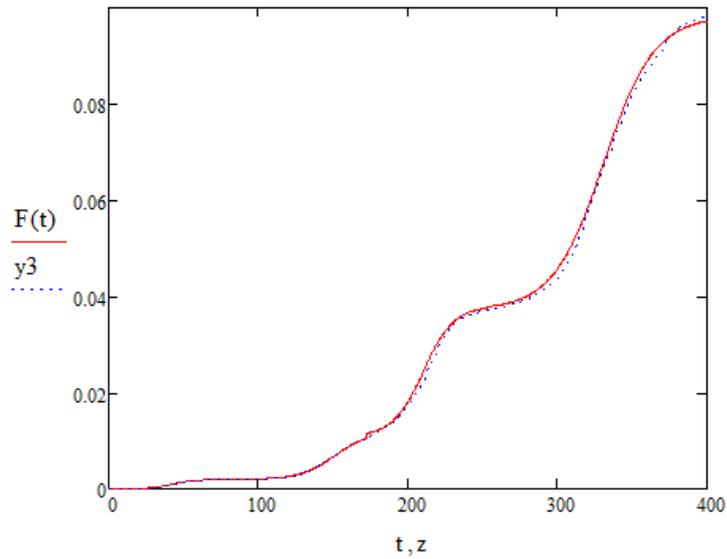

Fig.9a. Logistic curve fitting of Israel Covid-19 chance of infection (x axis: days since the first case, y axis: dot line - empirical data, solid line – function fit)

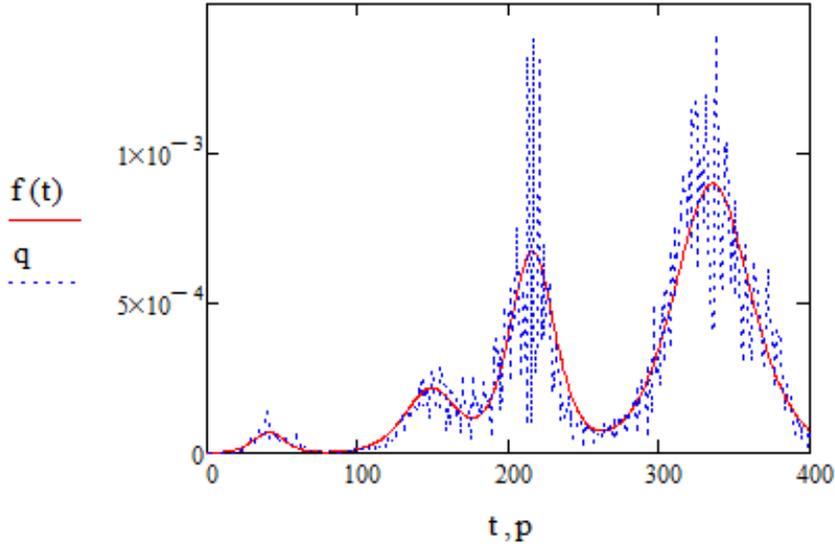

Fig.9b. Israel Covid-19 new cases chance of infection four wave approximation (x axis: days since the first case, y axis: dot line - empirical data, solid line – function fit)

In order to test the model I compare the ratio of the amplitudes of subsequent waves to the amplitude of the first wave and the ratio of the time of the peak of subsequent waves to the time of the peak of the first wave for the set of eight countries. The results are summarized in Table 2. The values $A_i$, $B_i$, $C_i$, $T_i$ are obtained from empirical data fit (Table 1)

Table 2 The ratio of the amplitudes of subsequent waves to the amplitude of the first wave and the ratio of the time of the peak of subsequent waves to the time of the peak of the first wave

| | country | time ratios | amplitude ratios |
|---|---|---|---|

| | | $T_i/T_1$ | $(A_i \cdot C_i)/(A_1 \cdot C_1)$ |
|---|---|---|---|
| | USA | 2.05 | 1.86 |
| | USA | 3.66 | 5.89 |
| | Canada | 4.37 | 4.00 |
| | Russia | 2.73 | 2.73 |
| | UK | 4.84 | 5.00 |
| | UK | 6.22 | 11.00 |
| | Belgium | 6.43 | 6.34 |
| | Finland | 3.3 | 2.83 |
| | Finland | 4.3 | 4.87 |
| | Japan | 3.01 | 2.48 |
| | Japan | 5.43 | 9.07 |
| | Israel | 3.77 | 3.14 |
| | Israel | 5.35 | 9.80 |
| | Israel | 8.39 | 13.20 |

One can mention the proportionality between the ratios of amplitudes and ratios of time. According to the model predictions (Eq. 7), time ratios should exactly equal amplitude ratios. In case the model predictions hold there should be significant correlation between sets of these quantities. And really in the above example, one obtains substantial linear correlation between these sets. The Pearson correlation coefficient is $r = .898$.

## 3.    Conclusion

Here is proposed a model which describes evolution of a TH system as a result of information processes within the system. The major finding is that redundancy density function can be presented as the solution of non-linear evolutionary equation and evolves in recognizable patterns so that summary redundancy growth can be mapped as overlapping logistic curves.

Logistic functions are widely implemented in different spheres, such as economics (diffusion of innovation), linguistics (language change), medicine (modeling of pandemic) etc. to describe the evolution of corresponding systems. The results obtained in the framework of the model are compared with empirically observed statistical data relating to Covid-19 epidemic spread. Relevant describing infectious disease model, such as SIR model, can be expressed in terms of TH model. The results suggest comparatively good agreement between model predictions and real data, which means that the information processes in the system can be considered as drivers of system evolution. Model predicted recognizable patterns can be used to forecast future system evolution, which may also be of interest for practitioners and policy makers.

Thus, Shannon's information theory can be embedded in the theory of evolution of complex self-organizing systems described in the framework of the TH model.

The subject of future research includes the study of the applicability of this model to other areas besides those considered here and the extension of the theory to the case of higher order helices.

Acknowledgment

The author is grateful to Loet Leydesdorff for valuable comments on an earlier draft of this paper.

The paper was prepared within the framework of the Basic Research Program at the National Research University Higher School of Economics (NRU HSE) and with support of the Russian Academic Excellence Project '5-100'.

**Appendix A**

Shannon informational entropy for temporal cyclic systems can be written as:

$$H = -\sum_{i=1}^{S} p_i \log p_i \qquad (A1)$$

Dubois showed (2019) that taking into account temporal cyclic systems:

$$H = H(t) = -\sum_{i=1}^{S} p_i(t) \log p_i(t) \qquad (A2)$$

with entropy and normalization conditions:

$$\frac{1}{T}\int_0^T \sum_{i=1}^{S} p_i(t)dt = 1, \; H_0 = \frac{1}{T}\int_0^T H(t)dt \qquad (A3)$$

in case $S = 2$ one obtains harmonic oscillator equation:

$$\begin{cases} \frac{dp_1}{dt} = -\frac{F}{p_{1,0}}\left(p_1 - p_{1,0}\right) \\ \frac{dp_2}{dt} = \frac{F}{p_{2,0}}\left(p_2 - p_{2,0}\right) \end{cases} \qquad (A4)$$

where $p_{i0} = \frac{1}{T}\int_0^T p_i(t)dt$ and $F$ is any function of $p_i, t$:

Following Dubois one can define the state of reference:

$$I_0 = -\sum_{i=1}^{S} p_{i,0} \log p_{i,0} \qquad (A5)$$

and develop informational entropy $H$ in Taylor's series around the reference state:

$$H = I_0 - \sum_{i=1}^{S}[(\log p_{i,0} + 1)(p_i - p_{i,0}) + \frac{(p_i - p_{i,0})^2}{2p_{i,0}} + \cdots O((p_i - p_{i,0})^3] \qquad (A6)$$

Substituting the expression (A5) into equation (A6) and neglecting the terms beyond the second degree one obtains:

$$H = -\sum_{i=1}^{S}[p_i \log p_{i,0} + (p_i - p_{i,0})] + D^* \qquad (A7)$$

where

$$D^* = \sum_{i=1}^{S}\left(\frac{(p_i - p_{i,0})^2}{2p_{i,0}}\right) \qquad (A8)$$

The condition for non-asymptotic stability of cyclic system is:

$$\frac{dD^*}{dt} = 0 \tag{A9}$$

Let $S = 2N$ then one of possible solution of Eq. (A7) is:

$$\begin{cases} \frac{dp_{j-1}}{dt} = -\frac{\gamma}{p_{j,0}}\left(p_j - p_{j,0}\right) \\ \frac{dp_j}{dt} = \frac{\gamma}{p_{j-1,0}}\left(p_{j-1} - p_{j-1,0}\right) \end{cases} \tag{A10}$$

$j = 2, 4, \ldots 2N$. Upon differentiating system (A9) by time we obtain:

$$\begin{cases} \frac{d^2 p_j}{dt^2} = \frac{\gamma^2}{p_{j,0}p_{j-1,0}}\left(p_j - p_{j,0}\right) \\ \frac{d^2 p_{j-1}}{dt^2} = \frac{\gamma^2}{p_{j,0}p_{j-1,0}}\left(p_{j-1} - p_{j-1,0}\right) \end{cases} \tag{A11}$$

The function $D^*$ corresponds to non-linear residue in (A6) which is a truncated version of (A5).

Using non-truncated equations (A5) we obtain:

$$\begin{cases} \frac{d^2 p_j}{dt^2} = \frac{\gamma^2}{p_{jo}p_{j-1,0}}\left(p_j - p_{j,0}\right) + C_j \\ \frac{d^2 p_{j-1}}{dt^2} = \frac{\gamma^2}{p_{j,0}p_{j-1,0}}\left(p_{j-1} - p_{j-1,0}\right) + C_{j-1} \end{cases} \tag{A12}$$

where $C_j = O\left(p_j - p_{j,0}\right)^3_{tt}$. When $p_j$ are smaller than $p_{j-1}$, in order to keep the same order of magnitude one can drop the terms beyond the second degree for the variable $p_j$ and the terms beyond the third degree for the variable $p_{j-1}$. In a similar manner this leads to the function $D^{**}$ defined analogously to $D^*$:

$$D^{**} = \sum_j^S \frac{(p_{j-1} - p_{j-1,0})^2}{2p_{j-1,0}} + \frac{(p_j - p_{j,0})^2}{2p_{j,0}} - \frac{(p_j - p_{j,0})^3}{6p_{j,0}^2} \tag{A13}$$

Differentiating $D^{**}$ by time and equating to zero $\frac{dD^{**}}{dt} = 0$ gives:

$$\begin{cases} \frac{dp_{j-1}}{dt} = \frac{\gamma}{p_{j,0}}\left(p_j - p_{j,0}\right) - \frac{\gamma}{2p_{j,0}{}^2}\left(p_j - p_{j,0}\right)^2 \\ \qquad \frac{dp_j}{dt} = -\frac{\gamma}{p_{j-1,0}}\left(p_{j-1} - p_{j-1,0}\right) \end{cases} \qquad (A14)$$

In analogy with (A10) for non-truncated version system (A13) yields an equation for non-harmonic oscillator:

$$\frac{d^2 p_j}{dt^2} = -\frac{\gamma^2}{p_{j,0}p_{j-1,0}}\left(p_j - p_{j,0}\right) + \frac{\gamma^2}{p_{j,0}{}^2 p_{j-1,0}}\left(p_j - p_{j,0}\right)^2 + C_j' \qquad (A15)$$

## Appendix B

Redundancy (Eq. 1) is a balance between two dynamics - evolutionary self-organization and historical organization, (Leydesdorff, 2010) or, in other words, between recursion on a previous state along the historical axis as opposed to meaning provided to the events from the perspective of hindsight (Dubois, 1998). Redundancy dynamics drives corresponding probabilities dynamics with recursive and incursive perspectives. Provided that probabilities oscillate in non-harmonic mode (Eq. A15) one can write:

$$\frac{d^2 p_j}{dt^2} = -\frac{\gamma^2}{p_{j,0} p_{j-1,0}} \left(p_j{}^- - p_{j,0}{}^-\right) + \frac{\gamma^2}{p_{j,0}{}^2 p_{j-1,0}} \left(p_j{}^- - p_{j0}{}^-\right)^2 + \frac{\gamma^2}{p_{j,0} p_{j-1,0}} \left(p_j{}^+ - p_{j,0}{}^+\right) -$$
$$\frac{\gamma^2}{p_{j,0}{}^2 p_{j-1,0}} (p_j{}^+ - p_{j,0}{}^+)^2 + C_j' + C_j'' \qquad (B1)$$

$p_j{}^-$ and $p_j{}^+$ are defined with respect to past and future states. Then Eq. (A3) using trapezoidal rule can be written as:

$$p_{j,0}{}^- = \frac{1}{2}(p_j{}^- + p_j); \ p_{j,0}{}^+ = \frac{1}{2}(p_j{}^+ + p_j); \ \text{so that} \ p_j{}^- - p_{j,0}{}^- = \frac{1}{2}(p_j - p_j{}^-); \ p_j{}^+ - p_{j,0}{}^+ = \frac{1}{2}(p_j{}^+ - p_j)$$

Developing $p_j{}^+$ and $p_j{}^-$ in Taylor's series in the state space:[6]

$$\begin{aligned} p_j{}^+ &= p_j + p_j' h + \frac{1}{2} p_j'' h^2 + \frac{1}{6} p_j''' h^3 + \frac{1}{24} p_j'''' h^4 + \cdots \\ p_j{}^- &= p_j - p_j' h + \frac{1}{2} p_j'' h^2 - \frac{1}{6} p_j''' h^3 + \frac{1}{24} p_j'''' h^4 + \cdots \end{aligned} \qquad (B2)$$

and keeping the terms up to $h^4$ order of magnitude one obtains (Fermi, Pasta, Ulam, 1955):

$$\frac{1}{k} p_{j_{tt}} = p_j'' h^2 + 2\alpha p_j' p_j'' h^3 + \frac{1}{12} p_j'''' h^4 + O(h^5) \qquad (B3)$$

$$k = \frac{\gamma^2}{p_{j,0} p_{j-1,0}}; \ \alpha = \frac{1}{p_{j-1,0}}; \ C_1 = \frac{1}{k}(C_j' + C_j'')$$

Setting further: $w = \sqrt{k}; \ \tau = wt; \ y = {}^x\!/_h; \ \varepsilon = 2\alpha$ one can rewrite (B3) in the form:

$$-p_{j_{tt}} + p_{j_{yy}} + \varepsilon p_{j_y} p_{j_{yy}} + \frac{1}{12} p_{j_{yyyy}} + C_1 = 0 \qquad (B4)$$

---

[6] The state space is presented by x axis

passing to the moving frame: $X = y - \tau$, rescaling time variable $T = \frac{\varepsilon}{2}\tau$, and keeping the terms up to the first order of $\varepsilon$ one brings the Eq. (B3) to the form:

$$\varepsilon\Sigma_{XT} + \varepsilon\Sigma_X\Sigma_{XX} + \frac{1}{12}\Sigma_{XXXX} + C_1 = 0 \qquad (B5)$$

Here $p_j = \Sigma(X, T)$. Defining further: $P = \Sigma_X$ and $\delta = \frac{1}{12\varepsilon}$ we obtain non-linear evolutionary equation:

$$P_T + PP_X + \delta P_{XXX} + C_1 = 0 \qquad (B6)$$

which corresponds to Korteweg de Vries (KdV) equation (Gibbon, 1985):

$$U_T + UU_X + \delta U_{XXX} = 0 \qquad (B7)$$

with additional term $C_1$. By substitution: $U \to \frac{1}{6}U$; $T \to \sqrt{\delta}T$; $X \to \sqrt{\delta}X$ Eq. (B7) is reduced to the form:

$$U_T + 6UU_X + U_{XXX} = 0 \qquad (B8)$$

Taking into account that equation (B8) possesses a soliton solution:

$$U = \frac{\kappa^2}{2}sech^2\left[\frac{k}{2}(X - \kappa^2 T)\right] \qquad (B9)$$

the final form of the solution of equation (B6) is:

$$P = \frac{\kappa^2}{12}sech^2\left[\frac{\kappa}{2}\left(\sqrt{\delta}(X - \kappa^2 T) + \frac{C_1}{2}T^2\right)\right] - C_1 T \qquad (B10)$$

the argument of the function $P$ at point $X = 0$ is 0 for values $T_0 = 0$ and $T_0' = \frac{2\kappa^2\sqrt{\delta}}{C_1}$

. For two solitons with amplitudes $A_1 = \frac{k_1^2}{12}$ and $A_2 = \frac{k_2^2}{12}$ the following condition is satisfied

$$A_1/A_2 = k_1/k_2$$